\documentclass[%
 aip,
 amsmath,amssymb,
 reprint,%
]{revtex4-2}

\usepackage{graphicx}
\usepackage{dcolumn}
\usepackage{bm}
\usepackage{hyperref}
\usepackage[utf8]{inputenc}
\usepackage[T1]{fontenc}
\usepackage{mathptmx}
\usepackage{etoolbox}

\usepackage{subfigure}
\usepackage{url}

\usepackage{color}
\usepackage{listings}

\hypersetup{hypertex=true,colorlinks=true,linkcolor=blue,anchorcolor=blue,citecolor=blue,urlcolor=blue}

\newcommand{\bs}[1]{\boldsymbol{#1}}

\makeatletter
\def\@email#1#2{%
 \endgroup
 \patchcmd{\titleblock@produce}
  {\frontmatter@RRAPformat}
  {\frontmatter@RRAPformat{\produce@RRAP{*#1\href{mailto:#2}{#2}}}\frontmatter@RRAPformat}
  {}{}
}%
\makeatother
\begin{document}

\preprint{AIP/123-QED}

\title{Development of a gyrokinetic-MHD energetic particle simulation code Part II: Linear simulations of Alfvén eigenmodes driven by energetic particles}

\author{Z.Y. Liu}
\affiliation{
Institute for Fusion Theory and Simulation and School of Physics, Zhejiang University, Hangzhou, Zhejiang 310027, People's Republic of China
}
\affiliation{
Zhejiang Lab, Hangzhou, Zhejiang 311121, People's Republic of China
}
\author{P.Y. Jiang}
\affiliation{
Institute for Fusion Theory and Simulation and School of Physics, Zhejiang University, Hangzhou, Zhejiang 310027, People's Republic of China
}
\author{S.Y. Liu}
\affiliation{
Institute for Fusion Theory and Simulation and School of Physics, Zhejiang University, Hangzhou, Zhejiang 310027, People's Republic of China
}
\author{L.L. Zhang}
\affiliation{
Institute for Fusion Theory and Simulation and School of Physics, Zhejiang University, Hangzhou, Zhejiang 310027, People's Republic of China
}
\author{G.Y. Fu}
\email{gyfu@zju.edu.cn}
\affiliation{
Institute for Fusion Theory and Simulation and School of Physics, Zhejiang University, Hangzhou, Zhejiang 310027, People's Republic of China
}

\date{\today}

\begin{abstract}
We have developed a hybrid code GMEC: Gyro-kinetic Magnetohydrodynamics (MHD) Energetic-particle Code that can numerically simulate energetic particle-driven Alfvén eigenmodes and energetic particle transport in tokamak plasmas. In order to resolve the Alfvén eigenmodes with high toroidal numbers effectively, the field-aligned coordinates and meshes are adopted. The extended MHD equations are solved with five-points finite difference method and fourth order Runge-Kutta method. The gyrokinetic equations are solved by particle-in-cell (PIC) method for the perturbed energetic particle pressures that are coupled into the MHD equations. Up to now, a simplified version of the hybrid code has been completed with several successful verifications including linear simulations of toroidal Alfvén eigenmodes and reversed shear Alfvén eigenmodes.
\end{abstract}

\maketitle

\section{Introduction}\label{sec1}
In burning plasmas, how to confine energetic particles (EP) including alpha particles is very important for future fusion reactors such as International Thermonuclear Experimental Reactor (ITER) and China Fusion Engineering Test Reactor (CFETR), since EPs are key for heating bulk plasma to attain ignition and high energy gain. Unfortunately, the EP characteristic orbit frequencies are comparable to the shear Alfvén wave frequencies in tokamaks. As a result, collective Alfvén eigenmode (AE) instabilities can be excited via wave-particle resonant interaction, leading to anomalous EP transport \cite{Chen2016, Gorelenkov2014, Heidbrink2008}. The EP transport can affect plasma equilibrium, MHD stability, bulk plasma transport, plasma heating and current drive. Due to the strong nonlinear coupling of EPs with bulk plasmas, it is necessary to develop advanced numerical simulation tools to investigate the underlying physics of EP-driven instabilities and EP transport.

There are different physical models to which different implementations can be applied in numerical simulations. For instance, perturbative eigenvalue code NOVA-K solves ideal MHD equations and evaluates the wave-particle interaction by employing the quadratic form with the perturbed distribution function of EPs obtained from the gyrokinetic (GK) equations \cite{Cheng19921, Gorelenkov19992802}. Similar codes include CAS3D-K \cite{Könies2008133}, AE3D-K \cite{Spong2010} and VENUS \cite{Cooper2011}. The non-perturbative methods can be classified into two main categories, GK model and kinetic-MHD hybrid model. For the former one, typical GK PIC initial value codes include GTC \cite{Lin19981835, Xiao2015}, GEM \cite{Chen2003463}, ORB5 \cite{Jolliet2007409} and EUTERPE \cite{Mishchenko2014}, and there are also Eulerian GK solvers such as GENE \cite{Görler20117053} and GYRO \cite{Candy2003545, Candy2003}. However, in order to study the AEs driven by EPs, the kinetic-MHD hybrid model seems to be more popular, because the hybrid approach requires significantly less simulation time while retaining the essential physics of wave-particle resonant interaction. The widely used hybrid codes include MEGA \cite{Todo19981321}, M3D-K \cite{Park19922033, Fu2006}, NIMROD \cite{Kim2008}, HMGC \cite{Briguglio19953711}, CLT-K \cite{Zhu2016} and M3D-C1-K \cite{Liu2022}. In these hybrid codes, bulk plasmas are described by reduced or full MHD equations in cylinder coordinates or toroidal flux coordinates. Spatial discretization methods include finite difference, finite element. Time advance can be explicit such as prediction-correction and Runge-Kutta methods, semi-implicit or fully implicit. Some codes use Fourier decompositions in poloidal and/or toroidal directions. EPs are taken into account by solving the drift-kinetic (DK) or GK equations via PIC methods, and corresponding distribution function can be calculated by full $f$ or $\delta f$ method \cite{Lin19955646}. And EPs are coupled into MHD equations through pressure or current coupling schemes \cite{Park19922033}.

We have developed a new hybrid code GMEC that can numerically simulate AEs driven by EPs in fusion plasmas. The ultimate goal of this work is to develop a highly efficient hybrid code that can be used to simulate alpha particle-driven Alfvén instabilities and alpha particle transport in burning plasmas. In the GMEC code, electrons are treated as a fluid, EPs and thermal ions are described by GK model, and the coupled set of hybrid equations is solved as an initial value problem. The magnetic-field-aligned coordinates and meshes are adopted to resolve the AEs with high toroidal numbers effectively. The MHD equations are solved with five-points finite difference method for spatial discretization in all three directions and fourth order Runge-Kutta method for time advance. The GK equations are solved by PIC method with multi-point gyro-averaging scheme. The perturbed distribution functions for EPs and thermal ions are solved by the $\delta f$ method and are used to compute the perturbed EP and thermal pressures that are coupled into the MHD equations. Up to now, a simplified version of the hybrid code has been completed with several verifications and benchmarks including (1) linear simulations of the $n=3$ toroidal Alfvén eigenmode (TAE) without FLR effect of EPs in an analytical circular equilibrium, and then in the corresponding VMEC \cite{Hirshman19833553} numerical equilibrium, where $n$ is the toroidal mode number. The results of GMEC agree well with the results of M3D-K code, (2) linear simulations of the $n=6$ TAE with and without FLR effect of EPs. The growth rate and mode structure of GMEC are in good agreement with previous results from other eigenvalue, kinetic and hybrid codes \cite{Könies2018}, (3) benchmark of the reversed shear Alfvén eigenmode (RSAE) observed in DIII-D experiments \cite{Collins2016}, the linear dispersion of $n=3-6$ modes agree reasonably with the results of the multi-code verification and validation simulations \cite{Taimourzadeh2019}.

This paper is the part II of the series papers named ‘Development of a gyrokinetic-MHD energetic particle simulation code’, with emphasis on the GK equation solver via PIC method and linear simulations of AEs driven by EPs. Details of the MHD equation solver and benchmarks of MHD modes are presented in the part I \cite{jiang2024development} of this series paper. The rest of this paper are organized as follows. In section \ref{sec2}, we give the basic equations and numerical approaches used in GMEC. GMEC code verifications and benchmarks are presented in section \ref{sec3}. Finally, a summary is given in section \ref{sec4}.

\section{Basic equations and numerical approaches}\label{sec2}
The basic equations refer to the reduced MHD equations and GK equations in section \ref{sec2.1}, and section \ref{sec2.2} introduces field-aligned coordinates and meshes adopted in GMEC. Section \ref{sec2.3} describes numerical approaches including spatial discretization and time advance, particles loading and gyro-averaging scheme, and parallelization strategies in high-performance computers. In section \ref{sec2.4} we describe the interface with analytical and VMEC numerical equilibriums in magnetic-field-aligned coordinates.

\subsection{Basic equations of kinetic-MHD hybrid model}\label{sec2.1}
For the first version of GMEC reported in this paper, MHD equations in GMEC hybrid model is chosen to be the same as Model B in part I of the series paper. For completeness, we repeat the reduced MHD equations presented in Part I as follows. The simplified vorticity equation is \cite{jiang2024development}
\begin{flalign}\label{eq1}
\begin{split}
    \frac{\partial}{\partial t}\delta \varpi = &\nabla\times(\delta A_{\parallel} \bs{b}_0)\cdot\nabla \frac{\mu_0 J_{\parallel}}{B} + B\bs{b}\cdot\nabla \frac{\mu_0 \delta J_{\parallel}}{B} \\
    &+ \frac{2\mu_0}{B} \bs{b}\times\bs{\kappa}\cdot\nabla(\delta P_b+\delta P_h),
\end{split}
\end{flalign}
where $\delta \varpi\equiv\nabla\cdot(1/v_A^2)\nabla \delta\varphi$ represents vorticity, $v_A\equiv B/\sqrt{\mu_0 n_i m_i}$ is the Alfvén velocity, $B$ is the equilibrium magnetic field, $n_i$ is the ion density, $m_i$ is the ion mass, $\delta\varphi$ is the perturbed electric potential, $\delta A_{\parallel}$ is the perturbed parallel magnetic potential, $\bs{b}$ is the unit vector along the equilibrium magnetic field, $\bs{\kappa}\equiv\bs{b}\cdot\nabla\bs{b}$ is the equilibrium magnetic curvature, $J_{\parallel}$ and $\delta J_{\parallel}$ are equilibrium and perturbed parallel current respectively. $\delta P_b$ is the perturbed bulk plasma pressure containing both thermal electrons and thermal ions, $\delta P_h$ is the perturbed EP pressure. The vorticity equation is closed by using the following relations
\begin{flalign}\label{eq2}
\begin{split}
    \delta J_{\parallel}=-\frac{1}{\mu_0 B}\nabla\cdot\left(B^2\nabla_{\perp}\frac{\delta A_{\parallel}}{B}\right),
\end{split}
\end{flalign}
\begin{flalign}\label{eq3}
\begin{split}
    \frac{\partial}{\partial t}\delta A_{\parallel} = -\bs{b}\cdot\nabla\delta\varphi,
\end{split}
\end{flalign}
\begin{flalign}\label{eq4}
\begin{split}
    \frac{\partial}{\partial t}\delta P_b = -\frac{1}{B}\bs{b}\times\nabla\delta\varphi\cdot\nabla P_b-\frac{2\Gamma P_b}{B}\bs{b}\times{\kappa}\cdot\nabla\delta\varphi,
\end{split}
\end{flalign}
where $P_b$ is equilibrium bulk plasma pressure, $\Gamma$ is the adiabatic coefficient and is set to be $5/3$.

We add numerical dissipation to $\delta \varpi$, by solving the diffusion equation of $\partial\delta\varpi/\partial t=D_{\delta\varpi}\nabla^2_{\perp}\delta\varpi$. This implementation is used to suppress the numerical instability associated with the field-aligned coordinates, which can also be stabilized by the shifted metric coordinates \cite{Scott2001447}, as shown in part I of the series paper. The shifted metric coordinates are not used in this work. Application of the shifted metric coordinates to the GK equation solver is left as future work. We also add numerical smoothing to $\delta P_h$ by solving the diffusion equation $\partial\delta P_h/\partial t=D_{\delta P_h}\nabla^2_{\perp}\delta P_h$, which can smooth the perturbed pressure and suppress the numerical noises introduced by PIC method. The diffusion coefficients are selected to be small enough such that they affect little the physical growth rate and real frequency of AEs.

The GK equations of motion are \cite{Fu2006}
\begin{flalign}\label{eq5}
\begin{split}
    \frac{d\bs{X}}{d t} = \frac{1}{B^{**}}\left\{v_{\parallel}\bs{B}^*-\bs{b}\times\left[\langle\delta \bs{E}\rangle-\frac{\mu}{q_s}\nabla(B+\langle\delta B\rangle)\right]\right\},
\end{split}
\end{flalign}
\begin{flalign}\label{eq6}
\begin{split}
    m_s\frac{dv_{\parallel}}{d t} = \frac{q_s}{B^{**}}\bs{B}^*\cdot\left[\langle\delta \bs{E}\rangle-\frac{\mu}{q_s}\nabla(B+\langle\delta B\rangle)\right],
\end{split}
\end{flalign}
where $\bs{X}$ is the guiding-center position, $B^{**}\equiv \bs{B}^*\cdot \bs{b}$ and
\begin{flalign}\label{eq7}
\begin{split}
    \bs{B}^*=\bs{B}+\langle\delta \bs{B}\rangle+\frac{m_s}{q_s}v_{\parallel}\nabla\times\bs{b},
\end{split}
\end{flalign}
$\mu\equiv m_s v^2_{\perp}/2B$, $v_{\parallel}$ and $v_{\perp}$ are parallel and perpendicular velocity respectively. $q_s$ and $m_s$ denote particle charge and mass of species $s$, in this paper we only treat EPs kinetically, while multiple species including EPs and thermal ions will be considered simultaneously in future work. $\delta \bs{E}\equiv -\nabla\delta\varphi-(\partial \delta A_{\parallel}/\partial t)\bs{b}$ is the perturbed electric field, and $\delta \bs{B}\equiv \nabla\times(\delta A_{\parallel} \bs{b})$ is the perturbed magnetic field. $\langle ... \rangle$ represents gyro-averaging.

\subsection{Field-aligned coordinates and meshes}\label{sec2.2}
The magnetic-field-aligned coordinates takes full advantage of the characteristic mode structure of AEs that their wavelength along the magnetic field direction is much larger than the perpendicular counterpart, \textit{i.e.}, $k_{\parallel}\ll k_{\perp}$, where $k_{\parallel}$ and $k_{\perp}$ are parallel and perpendicular wavenumber respectively. As a result, we can use relatively fewer grids in the parallel direction to accelerate computation. Meanwhile, when obtaining $\delta\varphi$ from the vorticity $\delta \varpi\equiv\nabla\cdot(1/v_A^2)\nabla_{\perp} \delta\varphi$, this 3D Poisson equation can be reduced to a simpler 2D one by neglecting $\nabla_{\parallel}$ terms.

Based on magnetic flux coordinates $(\psi,\theta,\phi)$, where $\psi\equiv\psi_p/\psi_{pm}$ is the normalized poloidal flux, $\psi_{pm}$ is the poloidal flux at plasma edge, $\theta$ and $\phi$ are the poloidal and toroidal angles respectively, the magnetic-field-aligned coordinates are defined as
\begin{flalign}\label{eq8}
\begin{split}
    x=\frac{\psi-\psi_1}{\Delta\psi}, y=\theta, z=\phi-\int_0^{\theta}d\theta^{\prime}\nu(\psi,\theta^{\prime}),
\end{split}
\end{flalign}
where $\Delta\psi\equiv\psi_2-\psi_1$, $\psi_1$ and $\psi_2$ are the left and right edge of the preassigned simulation region. Notice that we have artificially excluded the area around the magnetic axis to get rid of the singularity at $\psi=0$. $\nu(\psi,\theta)\equiv\bs{b}\cdot\nabla\phi/\bs{b}\cdot\nabla\theta$ is the local field line pitch. In this paper, we make use of the straight field line flux coordinates, hence $\nu(\psi,\theta)=q(\psi)$ is the safety factor. Due to the periodicity in $\theta$ and $\phi$ directions, the perturbed fields, for example $\delta \varphi$, in the magnetic-field-aligned coordinates satisfy
\begin{flalign}\label{eq9}
\begin{split}
    \delta \varphi(x,y,z+2\pi)&=\delta \varphi(x,y,z),\\
    \delta \varphi(x,y+2\pi,z)&=\delta \varphi(x,y,z-2\pi q),
\end{split}
\end{flalign}
which is the so-called twist-shift boundary condition \cite{DUDSON20091467}.

\begin{figure}[htbp]
\centering
\includegraphics[width=0.45\textwidth]{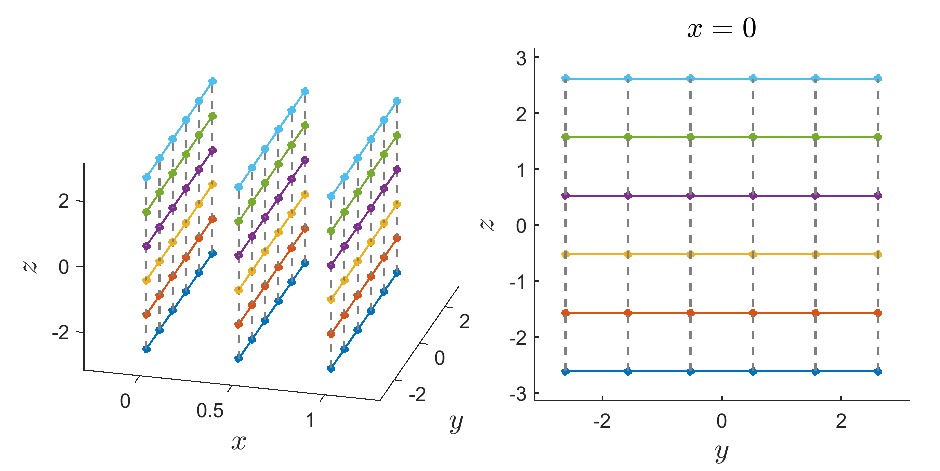}
\caption{Structured $3\times 6\times 6$ meshes in the $(x,y,z)$ coordinates, six magnetic field lines in the three magnetic surfaces are labeled by different colors.}
\label{fig1}
\end{figure}

We adopt 3D structured meshes with uniform grid sizes in the $(x,y,z)$ coordinates. For example, if the grid numbers are $n_x,n_y,n_z$, the discretized meshes range from $0$ to $1$ with an interval of $\Delta x=1/(n_x-1)$ in $x$ direction, from $-\pi+\Delta y/2$ to $\pi-\Delta y/2$ with an interval of $\Delta y=2\pi/n_y$ in $y$ direction, and from $-\pi+\Delta z/2$ to $\pi-\Delta z/2$ with an interval of $\Delta z=2\pi/n_z$ in $z$ direction. A structured $3\times 6\times 6$ meshes are illustrated in Figure \ref{fig1}, in which six magnetic field lines in the three magnetic surfaces are labeled by different colors.

\begin{figure}[htbp]
\centering
\includegraphics[width=0.45\textwidth]{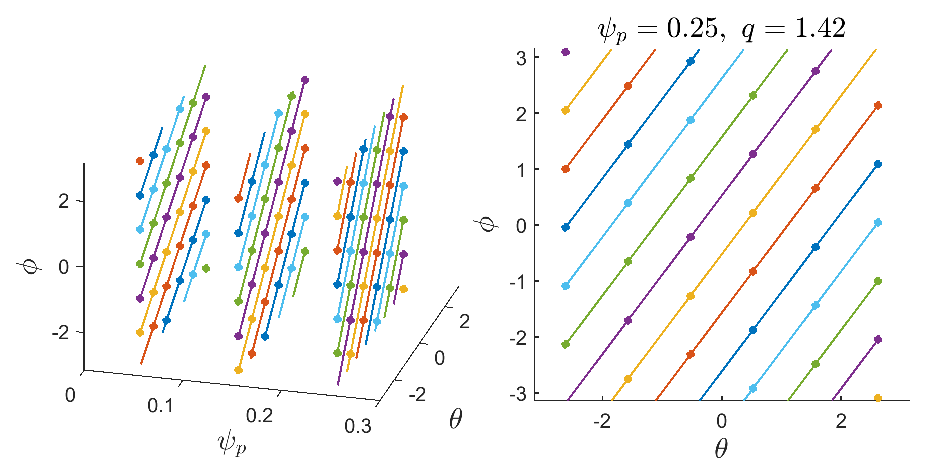}
\caption{Magnetic field lines in Figure \ref{fig1} plotted in $(\psi_p,\theta,\phi)$ coordinates.}
\label{fig2}
\end{figure}

In Figure \ref{fig2}, we convert these magnetic field lines in Figure \ref{fig1} into $(\psi_p,\theta,\phi)$ coordinates, where the $q$ profile is set to be $q=0.5+1.5(r/a_0 )^2$, $d\psi_p/dr=rB_0/q$, and use is made of the usual tokamak parameters with minor radius $a_0=0.6\mathrm{m}$, $B_0=2\mathrm{T}$. The magnetic surface of $\psi_p=0.25$ and $q=1.42$ is illustrated on the right in Figure \ref{fig2}. We can see that these magnetic field lines are able to cover the whole magnetic surface. This is a little different from the flux-tube coordinates \cite{Beer19952687}, in which only one magnetic field line is followed and the simulation is limited to a band region in the magnetic surface.

\subsection{Numerical approaches and parallel strategies}\label{sec2.3}
In the $(x,y,z)$ coordinates, Jacobi is $J\equiv(\nabla x\times\nabla y\cdot\nabla z)^{-1}$, the contravariant and covariant metric coefficients are $g^{ij}=\nabla u^i\cdot\nabla u^j$ and $g_{ij}=\bs{e}_i\cdot\bs{e}_j$ respectively. Here $u^i\equiv x,y,z$, $i=1,2,3$, $\bs{e}_i$ is the covariant basis vector corresponding to $u^i$. The equilibrium quantities including $q, n_i, P_b, B, J_{\parallel}$ and geometric coefficients of $J,g^{ij},g_{ij}$, and also their derivatives (usually first-order derivatives are enough, but second-order derivative of $B$ is also necessary) such as $\partial_i B$, $\partial_i \equiv\partial/\partial u^i$, are all precalculated on the mesh grids and saved into an equilibrium file.

With the notations mentioned above, the vector operators in the reduced MHD equations in section \ref{sec2.1} can be simplified as different combinations of the precalculated equilibrium quantities and differential operators of $\partial_i$ acting on the perturbed fields. When expanding the equations in $(x,y,z)$ coordinates, we get massive and complicated expressions, which are very difficult to code them manually. This problem has been handled by a system of symbolic operations to implement equations and load matrix automatically as described in part I \cite{jiang2024development} of the series paper. In this paper, five-points finite difference method up to fourth order precision has been adopted. For time advance, we use the fourth order Runge-Kutta method both for the reduced MHD equations and GK equations of motion.

The GK equations of motion in $(x,y,z)$ coordinates are
\begin{flalign}\label{eq10}
\begin{split}
    \frac{du^i}{dt}=\frac{1}{B^{**}}(v_{\parallel}B^{*i}-\hat{d}^i F),\ i=1,2,3,
\end{split}
\end{flalign}
\begin{flalign}\label{eq11}
\begin{split}
    \frac{dv_{\parallel}}{dt}=-\frac{q_s}{m_s}\left(\frac{\partial}{\partial t}\langle\delta A_{\parallel}\rangle+\frac{1}{B^{**}}\sum_i B^{*i}\partial_i F\right),
\end{split}
\end{flalign}
where $F\equiv\langle\delta\varphi\rangle+(\mu/q_s)(B+\langle\delta B\rangle)$, $B^{**}=(\Delta \psi_p/JB)\sum_i g_{yi}B^{*i}$,
\begin{flalign}\label{eq12}
\begin{split}
    \bs{B}^{*}=\sum_i B^{*i}\bs{e}_i=\frac{\Delta\psi_p}{J}\bs{e}_y+(\bs{c}+\hat{\bs{d}})\langle\delta A_{\parallel}\rangle+\frac{m_s}{q_s}\bs{c}v_{\parallel},
\end{split}
\end{flalign}
and $\bs{c}\equiv\nabla\times\bs{b}$ with contravariant components $c^i=(\Delta \psi_p/J)\sum_{j,k}\epsilon^{ijk}\partial_j (g_{yk}/JB)$, the operator $\hat{\bs{d}}$ is defined as $\hat{\bs{d}}f\equiv-\bs{b}\times\nabla f$ with contravariant components $\hat{d}^i=(\Delta\psi_p/J^2B)\sum_{j,k}\epsilon^{ijk}g_{yk}\partial_j$. Here $\Delta\psi_p=\psi_{pm}\Delta\psi$, $\epsilon^{ijk}$ is the Levi-Civita symbol. The perturbed magnetic field strength can be approximated as $\delta B\approx\bs{b}\cdot\delta\bs{B}=(\Delta\psi_p/JB)\sum_i g_{yi}(c^i+\hat{d}^i)\delta A_{\parallel}$.

Pushing GK equation of motion needs equilibrium and perturbed fields, and also their derivatives at current guiding-center position. These quantities are obtained by linear interpolation of values at adjacent eight mesh grids, on which equilibrium quantities and their derivatives have been precalculated and read from the equilibrium file at the beginning of GMEC run, and the perturbed quantities are updated using the MHD equation solver at each time step. This implementation is also known as gathering in PIC method. The time step $\Delta t$ adopted in MHD and GK equation solver is set to be the same value and around $0.01\tau_A$, where $\tau_A=1/\omega_A$, $\omega_A=v_{A0}/R_0$ is the Alfvén frequency, $v_{A0}$ is the Alfvén velocity at magnetic axis, $R_0$ is the major radius of tokamak. Normalizations have also been unified, the length is normalized by $L_0\equiv\sqrt{\Delta\psi_p/B_0}$, velocity is normalized by $v_{A0}$, hence the time is normalized by $L_0/v_{A0}$. Normalizations of other quantities can be determined by $L_0$ and $v_{A0}$. More details of normalization are given in Part I \cite{jiang2024development} of this series paper.

Neglecting $\langle ... \rangle$ in the GK equations, we solve GK equations without FLR effect. In GMEC code, multi-point gyro-averaging scheme \cite{Xiao2015, Lin19955646} is adopted to solve the GK equations with FLR effect. In order to be applicable to the non-orthogonal $(x,y,z)$ coordinates, an improved gyro-averaging strategy which are originally designed for GTC code to overcome the problem of strongly shaped plasmas \cite{Duan2022} has been employed. We make the same assumption as in GTC that the gyro-planes are approximated the poloidal planes with equal $\phi$ in tokamaks. The specific steps are as follows, (1) for the guiding-center position $(x_i,y_i,z_i)$, we convert it into the cylinder coordinate as $(R_i,Z_i,\phi_i)$, the values of $q$ and its derivative at this position are denoted as $q_0$ and $q_0^{\prime}$, (2) we find the two points of $(x_1,y_1,z_1 )=(x_{i}+\Delta x,y_i,z_i)$ and $(x_2,y_2,z_2 )=(x_i,y_i+\Delta y,z_i)$ with the corresponding cylinder coordinates $(R_{1,2},Z_{1,2},\phi_{1,2})$, and their distances to the guiding-center computed as $d_{1,2}$ respectively, (3) the Larmor radius of the particle is defined as $\rho_c=m_s v_{\perp,i}/q_s B$, where $B$ is evaluated at the guiding-center position, we define the two lengths as $\delta x=(\rho_c/d_1)\Delta x$ and $\delta y=(\rho_c/d_2)\Delta y$. The positions of $N$ gyro-averaging points $(x_i+\delta x_j,y_i+\delta y_j,z_i+\delta z_j)$, are given by
\begin{flalign}\label{eq13}
\begin{split}
    \delta x_j&=\sin(\phi_j+\phi_0)\frac{\delta x}{\sin\alpha},\\
    \delta y_j&=\sin(\phi_j+\phi_0-\alpha)\frac{\delta y}{\sin\alpha},
\end{split}
\end{flalign}
where $\phi_j=2\pi j/N$, $j=1,2,...,N$, $\phi_0$ is an arbitrary phase and we set it to be $\alpha/2$, $\alpha$ is the angle between contravariant basis vectors in $x,y$ directions,
\begin{flalign}\label{eq14}
\begin{split}
    \cos\alpha=\frac{\nabla x\cdot\nabla y}{|\nabla x||\nabla y|}=\frac{g^{xy}}{\sqrt{g^{xx}g^{yy}}},
\end{split}
\end{flalign}
and $\delta z_j$ should satisfy the equal $\phi$ condition $\delta(z+qy)=0$, which is approximated as $\delta z_j=-q_0\delta y_j-\Delta\psi q_0^{\prime}y_0 \delta x_j$. In the simulations in section \ref{sec3}, $N$ is set to be $4$.

Particles are loaded uniformly in the phase space via Monte Carlo methods. In order to ensure a uniform distribution in 3D physical space, we use the Jacobi as weight to generate particles. The first step is to produce a set of random coordinates $(\tilde{x},\tilde{y},\tilde{z})$ and corresponding Jacobi $\tilde{J}$, then we produce another random number $\tilde{r}\in [0,1]$, if $\tilde{r}\geq\tilde{J}/J_{\mathrm{max}}$, where $J_{\mathrm{max}}$ is the maximum Jacobi in the simulation region, a particle is loaded at the position of $(\tilde{x},\tilde{y},\tilde{z})$, otherwise we produce a new set of random coordinates until the condition mentioned above is met. Furthermore, a uniform distribution in $(v_{\parallel},v_{\perp}^2)$ space can be generated in a similar way.

The perturbed distribution function is calculated by the $\delta f$ method. In the $\delta f$ method, the total distribution function is split into an equilibrium one $f_0$ and a perturbed one $\delta f$. We define a particle weight $w\equiv\delta f/g$, where $g$ is the distribution of loaded particles (or markers). The evolution equation for the weight is \cite{Fu2006}
\begin{flalign}\label{eq15}
\begin{split}
    \frac{dw}{dt}=-\left(\frac{f}{g}-w\right)\frac{1}{f_0}\frac{df_0}{dt}.
\end{split}
\end{flalign}
The equilibrium distribution is expressed as a function of the constants of motion, $f_0=f_0(P_\phi,E,\mu)$, then
\begin{flalign}\label{eq16}
\begin{split}
    \frac{1}{f_0}\frac{df_0}{dt}=\left(\frac{1}{f_0}\frac{\partial f_0}{\partial P_{\phi}}\right)\frac{dP_{\phi}}{dt}+\left(\frac{1}{f_0}\frac{\partial f_0}{\partial E}\right)\frac{dE}{dt},
\end{split}
\end{flalign}
where $dP_{\phi}/dt$ and $dE/dt$ can be deduced from the GK equations. Here, $P_{\phi}$ is the toroidal angular momentum. In $(x,y,z)$ coordinates, it can be expressed as
\begin{flalign}\label{eq17}
\begin{split}
    P_{\phi}=\Delta\psi_p m_s v_{\parallel}\frac{g_{yz}}{JB}-q_s\psi_p.
\end{split}
\end{flalign}
In GMEC code, the equilibrium distribution $f_0$ can be either a slowing-down or a Maxwellian. The numerical expressions of $g$ and $\delta f$ are respectively
\begin{flalign}\label{eq18}
\begin{split}
    g=\frac{1}{N_p}\sum_i^{N_p}\left[\frac{1}{J}\delta(x-x_i)\delta(y-y_i)\delta(z-z_i)\right.\\
    \left.\frac{1}{2\pi v_{\perp}}\delta(v_{\parallel}-v_{\parallel,i})\delta(v_{\perp}-v_{\perp,i})\right],
\end{split}
\end{flalign}
\begin{flalign}\label{eq19}
\begin{split}
    \delta f=\frac{1}{N_p}\sum_i^{N_p}\left[w_i\frac{1}{J}\delta(x-x_i)\delta(y-y_i)\delta(z-z_i)\right.\\
    \left.\frac{1}{2\pi v_{\perp}}\delta(v_{\parallel}-v_{\parallel,i})\delta(v_{\perp}-v_{\perp,i})\right],
\end{split}
\end{flalign}
where $N_p$ is the number of loaded particles, and $i$ represents the $i^{\mathrm{th}}$ particle. The perturbed pressures of species $s$ in parallel and perpendicular directions are respectively
\begin{flalign}\label{eq20}
\begin{split}
    \delta P_{\parallel,s}=\frac{1}{N_p}\sum_i^{N_p}m_s v_{\parallel,i}^2w_i \frac{1}{J}\delta(x-x_i)\delta(y-y_i)\delta(z-z_i),
\end{split}
\end{flalign}
\begin{flalign}\label{eq21}
\begin{split}
    \delta P_{\perp,s}=\frac{1}{N_p}\sum_i^{N_p}\frac{1}{2}m_s v_{\perp,i}^2w_i \frac{1}{J}\delta(x-x_i)\delta(y-y_i)\delta(z-z_i).
\end{split}
\end{flalign}
The total perturbed pressure is $\delta P_s=(\delta P_{\parallel,s}+\delta P_{\perp,s})/2$. In practical calculations, when solving DK equations, the perturbed pressures at the guiding-center position are scattered to the adjacent eight mesh grids, which is the inverse process of gathering implementation. While for GK equations, a pull-back transformation needs to be performed on the perturbed pressures by substituting the following relation
\begin{flalign}\label{eq22}
\begin{split}
    &\delta(x-x_i)\delta(y-y_i)\delta(z-z_i)\rightarrow\\
    &\frac{1}{N}\sum_j^N\delta(x-x_i-\delta x_j)\delta(y-y_i-\delta y_j)\delta(z-z_i-\delta z_j),
\end{split}
\end{flalign}
into equations \ref{eq20} and \ref{eq21}.

GMEC has been parallelized using both multi-process and multi-threading. Message Passing Interface (MPI) is used to parallelize the tasks in different Central Processing Units (CPUs). The Intel library of Thread Building Block (TBB) \cite{Pheatt2008} is a sheared memory parallel method which is used to parallelize the tasks with multiple threads within one CPU. In GMEC code, the domain decomposition is only employed in $y$ direction, and the Poisson equation which has been reduced to a 2D problem in the $x,z$ plane is solved in each CPU by the PARDISO solver with TBB parallel speedup \cite{jiang2024development}. Each CPU only needs to handle the mesh grids and particles in its own domain with the aid of ghost grids in $y$ direction. MPI communication is used to update the information on ghost grids and particles that move out of their local domain. Notice that the boundary condition in equation \ref{eq9} should be taken into account for the updates around $y,z=\pm\pi$. Particles that move into regions of $x<0$ or $x>1$ are thrown out at present for simplicity. More self-consistent treatment is left as future work.

\subsection{Analytical and numerical equilibria}\label{sec2.4}
As shown in preceding sections, the equilibrium quantities including their derivatives are precalculated on the mesh grids and saved into an equilibrium file. The equilibrium information is all contained in these quantities, and separated from equation solvers. This scheme can be applicable to any curvilinear coordinates, and even generalized to simulations of stellarators in which equilibriums are 3D instead. In the paper, we focus on axisymmetric tokamaks.

First, an analytical equilibrium of concentric circles is derived as the footstone for the numerical one. The equilibrium is defined as
\begin{flalign}\label{eq23}
\begin{split}
    R=R_0+r\cos\theta_s,Z=r\sin\theta_s,\phi=\phi_s,
\end{split}
\end{flalign}
where $(r,\theta_s,\phi_s)$ are the geometric radius, poloidal and toroidal angles, and are related to the straight field line coordinates $(\psi,\theta,\phi)$ by $r=r(\psi)$, $\theta_s=\theta_s(\psi,\theta)$. The former function is determined by $d\psi_p/dr=rB_0/q$ for a given $q$ profile, while the latter function is determined by the condition that $J_{\psi,\theta,\phi}/R^2$ is only a function of $\psi$. These straight field line coordinates are also known as PEST coordinates \cite{White20131}. We use PEST coordinates in this paper, other straight field line coordinates are also applicable to GMEC, for example, the MHD solver in part I of the series paper has adopted Boozer coordinates. The determinant of the coordinate transformation from $(R,Z,\phi)$ to $(\psi,\theta,\phi)$ is $D=rr_{\psi}\theta_{s,\theta}$, where the subscript of $\psi,\theta$ denotes the partial derivative with respect to $\psi,\theta$ respectively. The Jacobi is $J_{\psi,\theta,\phi}=DR$, substituting it into the condition above leads to the differential equation $\theta_{s,\theta}=R/R_0$. Correct up to $O(\epsilon^2)$, $\epsilon\equiv r/R_0$, the solution is given approximated as
\begin{flalign}\label{eq24}
\begin{split}
    \theta_s=\theta+\frac{\epsilon\sin\theta+(\epsilon^2/4)\sin 2\theta}{1-\epsilon^2/2}.
\end{split}
\end{flalign}
The magnetic field in $(\psi,\theta,\phi)$ coordinates is $\bs{B}=\nabla\psi_p\times\nabla(q\theta-\phi)$, the strength of $\bs{B}$ is
\begin{flalign}\label{eq25}
\begin{split}
    B=\frac{\psi_{pm}}{D}\sqrt{q^2+\epsilon^2\theta^2_{s,\theta}},
\end{split}
\end{flalign}
and the contravariant metric coefficients $g^{\psi\psi},g^{\theta\theta},g^{\psi\theta},g^{\phi\phi}$ in $(\psi,\theta,\phi)$ coordinates can be calculated straightforwardly. Use is made of equation \ref{eq8} to calculate the Jacobi and contravariant metric coefficients in the magnetic-field-aligned coordinates as $J=\Delta\psi J_{\psi,\theta,\phi}$,
\begin{flalign}\label{eq26}
\begin{split}
    &g^{xx}=\frac{g^{\psi\psi}}{\Delta\psi^2},g^{yy}=g^{\theta\theta},g^{xy}=\frac{g^{\psi\theta}}{\Delta\psi},\\
    &g^{zz}=q^{\prime 2}\theta^2 g^{\psi\psi}+q^2g^{\theta\theta}+g^{\phi\phi}+2qq^{\prime}\theta g^{\psi\theta},\\
    &g^{yz}=-q^{\prime}\theta g^{\psi\theta}-q g^{\theta\theta},g^{xz}=-\frac{1}{\Delta\psi}(q^{\prime}\theta g^{\psi\psi}+q g^{\psi\theta}),
\end{split}
\end{flalign}
and the covariant metric coefficients can be calculated from contravariant ones by
\begin{flalign}\label{eq27}
\begin{split}
    g_{il}=J^2\sum_{j,k,m,n}\epsilon_{ijk}\epsilon_{lmn}(g^{jm}g^{kn}-g^{jn}g^{km}),
\end{split}
\end{flalign}
where $q^{\prime}\equiv\partial q/\partial\psi$ and $\epsilon_{ijk}$ is the Levi-Civita symbol.

The numerical equilibrium is generated from VMEC code \cite{Hirshman19833553} in this paper, and can also be generated from other equilibrium code such as DESC \cite{Dudt2020}, which has been described in Appendix C and D in part I of the series paper. The Fourier coefficients of $R(\psi,\theta)=\sum_m R^c_m(\psi)\cos m\theta$, $Z(\psi,\theta)=\sum_m Z^s_m(\psi)\sin m\theta$, $m=0,1,2,...$ can be obtained from the VMEC output. The Jacobi of the coordinate transformation from $(R,Z,\phi)$ to $(\psi,\theta,\phi)$ is $J_{\psi,\theta,\phi}=(R_{\psi}Z_{\theta}-R_{\theta}Z_{\psi})R$, and the contravariant metric coefficients are
\begin{flalign}\label{eq28}
\begin{split}
    &g^{\psi\psi}=\frac{R^2}{J^2_{\psi,\theta,\phi}}(R^2_{\theta}+Z^2_{\theta}),g^{\theta\theta}=\frac{R^2}{J^2_{\psi,\theta,\phi}}(R^2_{\psi}+Z^2_{\psi}),\\
    &g^{\psi\theta}=-\frac{R^2}{J^2_{\psi,\theta,\phi}}(R_{\psi}R_{\theta}+Z_{\psi}Z_{\theta}),g^{\phi\phi}=\frac{1}{R^2}.
\end{split}
\end{flalign}
Substituting it into equations \ref{eq26} and \ref{eq27} we obtain the contravariant and covariant metric coefficients in the magnetic-field-aligned coordinates.

\section{Code verifications and benchmarks}\label{sec3}
Code verifications and benchmarks of the GK equation solver and linear simulations of EP-driven AEs driven are presented in this section. Details of the MHD equation solver and benchmarks of MHD modes are presented in Part I of this series paper. Section \ref{sec3.1} gives verifications of single particle orbits in axisymmetric equilibrium magnetic field with and without a perturbed electromagnetic field. Linear simulations of a $n=3$ TAE are given in section \ref{sec3.2}, and benchmarks of a $n=6$ TAE and $n=3-6$ RSAEs are given in section \ref{sec3.3} and \ref{sec3.4} respectively.

\subsection{Particle orbit verifications}\label{sec3.1}
For single particle’s guiding-center motion in equilibrium magnetic field, when the particle’s energy $E$ and the inverse aspect ratio $\epsilon$ are small enough, there are analytical solutions for the orbit width, bounce/transit and processional frequencies \cite{White20131}. Here, we give the expressions correct up to $O(\epsilon^2)$. For trapped particles ($\kappa<1$), the orbit width is
\begin{flalign}\label{eq29}
\begin{split}
    \Delta r=\frac{4q}{q_s}\left(\frac{\epsilon\mu m_s}{B_0}\right)^{1/2}\frac{\kappa}{\epsilon(1-\epsilon)},
\end{split}
\end{flalign}
the bounce frequency is
\begin{flalign}\label{eq30}
\begin{split}
    \omega_b=\left(\frac{\epsilon\mu B_0}{m_s}\right)^{1/2}\frac{\pi}{2qR_0}\frac{1}{K(\kappa)+\epsilon(2E(\kappa)-K(\kappa))},
\end{split}
\end{flalign}
the toroidal processional frequency is
\begin{flalign}\label{eq31}
\begin{split}
    \omega_d=\frac{1}{q_sR_0}&\left[\frac{4\mu q \hat{s}}{r}\frac{E(\kappa)+(\kappa^2-1)K(\kappa)}{K(\kappa)}\right.\\
    &\left.+\frac{\mu q}{r}\frac{2E(\kappa)-K(\kappa)}{K(\kappa)}\right],
\end{split}
\end{flalign}
where $\kappa\equiv v_{\parallel}/\sqrt{4\epsilon\mu}$, $\hat{s}\equiv(r/q)(dq/dr)$ is the magnetic shear, $K(\kappa),E(\kappa)$ are the complete elliptic integral of the first and second kind respectively.

For passing particles ($\kappa>1$), the orbit width is
\begin{flalign}\label{eq32}
\begin{split}
    \Delta r=\frac{2q}{q_s}\left(\frac{\epsilon\mu m_s}{B_0}\right)^{1/2}\frac{1}{\epsilon}\left(\frac{\kappa}{1-\epsilon}-\frac{\sqrt{\kappa^2-1}}{1+\epsilon}\right),
\end{split}
\end{flalign}
the transit frequency is
\begin{flalign}\label{eq33}
\begin{split}
    \omega_t=\frac{\sqrt{\epsilon\mu B_0/m_s}(\pi\kappa/qR_0)}{K(1/\kappa)+\epsilon [(1-2\kappa^2)K(1/\kappa)+2\kappa^2E(1/\kappa)]}.
\end{split}
\end{flalign}

\begin{figure}[htbp]
\centering
\includegraphics[width=0.45\textwidth]{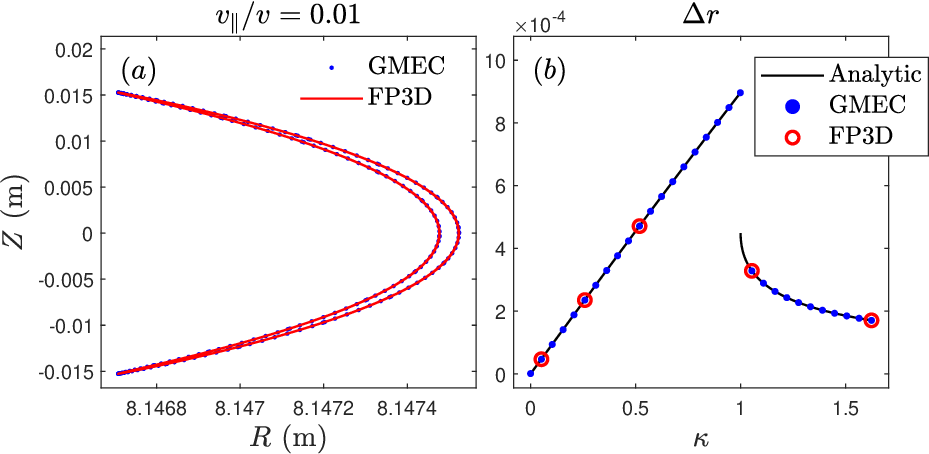}
\caption{(a) A trapped particle orbit of GMEC (blue points) and the result of another particle orbit code named FP3D (red line), (b) orbit width as function $\kappa$, analytic solution (black line) given by equations \ref{eq29} and \ref{eq32}, results of GMEC (blue points) and PF3D (red circles).}
\label{fig3}
\end{figure}

In order to obtain particle orbit verifications, the equilibrium and perturbed fields, and also their derivatives at current guiding-center position are evaluated by analytical equilibrium functions in equations \ref{eq26} and \ref{eq27} when pushing particle’s guiding-center motion equations. In Figure \ref{fig3}, a trapped particle orbit with the parameters of $E=1\mathrm{eV}$, $\epsilon=0.01844$, $R_0=8\mathrm{m}$, $a_0=0.6\mathrm{m}$, $B_0=2\mathrm{T}$, and $q=0.5+1.5(r/a_0 )^2$ is illustrated as blue points, which coincide with the result from another particle orbit solver FP3D \cite{Jiang2024} as shown by the red line. The analytical solutions of orbit width as function of $\kappa$ given by equations \ref{eq29} and \ref{eq32} are displayed in Figure \ref{fig3}(b), and results of GMEC and FP3D are shown by blue points and red circles respectively. The analytical and numerical results are in very good agreement.

\begin{figure}[htbp]
\centering
\includegraphics[width=0.45\textwidth]{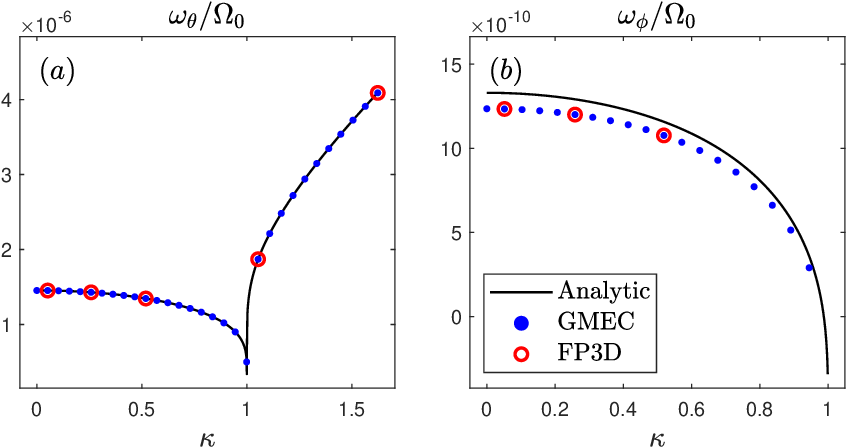}
\caption{(a) The bounce/transit and (b) processional frequencies as function of $\kappa$, analytic solution (black line) of $\omega_{\theta}$ given by equations \ref{eq30}, \ref{eq33} and $\omega_{\phi}$ given by equation \ref{eq31}, results of GMEC (blue points) and PF3D (red circles).}
\label{fig4}
\end{figure}

The analytical solutions of bounce/transit frequencies given by equations \ref{eq30} and \ref{eq33}, and processional drift frequency given by equation \ref{eq31} are also consistent with the numerical results of GMEC and FP3D as displayed in Figure \ref{fig4}, where $\Omega_0\equiv q_s B_0/m_s$ is the gyro-frequency.

\begin{figure}[htbp]
\centering
\includegraphics[width=0.45\textwidth]{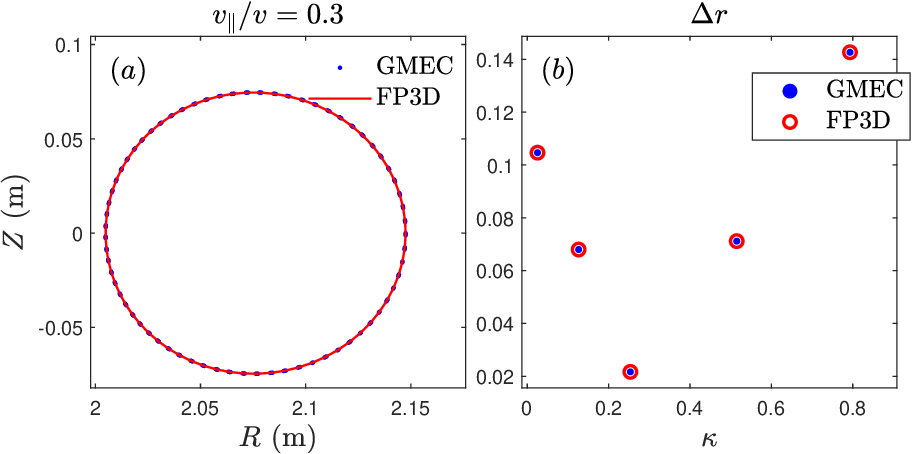}
\caption{(a) An EP orbit of GMEC (blue points) and the result of FP3D (red line), (b) orbit width as function $\kappa$, results of GMEC (blue points) and PF3D (red circles).}
\label{fig5}
\end{figure}

\begin{figure}[htbp]
\centering
\includegraphics[width=0.45\textwidth]{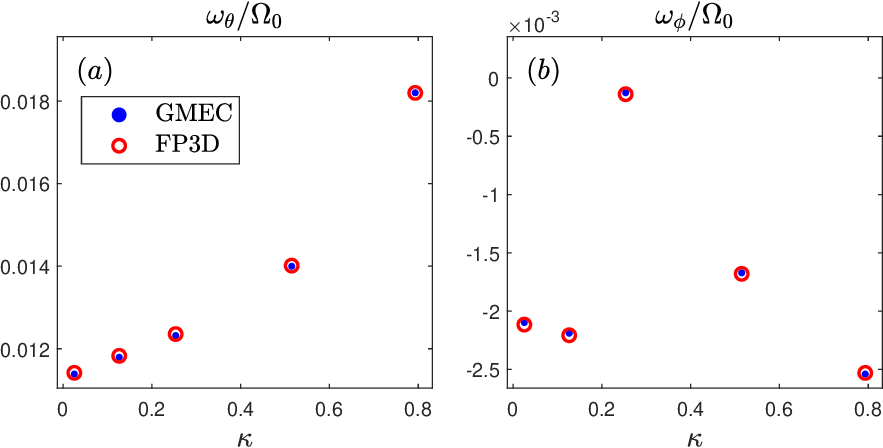}
\caption{Numerical solutions of (a) $\omega_{\theta}$ and (b) $\omega_{\phi}$ as function of $\kappa$ from results of GMEC (blue points) and PF3D (red circles).}
\label{fig6}
\end{figure}

For EPs, there are no analytical solutions for particles’ orbits. In Figure \ref{fig5}(a), an EP orbit from GMEC with the parameters of $E=1\mathrm{MeV}$, $\epsilon=0.07376$, $R_0=2\mathrm{m}$ (other parameters are the same as above), is illustrated as blue points. Particle orbit width as function of $\kappa$ is displayed in Figure \ref{fig5}(b). The numerical results of the two codes are almost identical. In Figure \ref{fig6} displayed are $\omega_{\theta}$ and $\omega_{\phi}$ as function of $\kappa$, there are little difference between the results of GMEC and FP3D.

The conservation of $E$ and $P_{\phi}$ has been carefully checked. For instance, relative variations of $E$ and $P_{\phi}$ in $10^4 \tau_A$ with $\Delta t=0.01 \tau_A$ are less than $10^{-7}$. These test cases give verifications of single particle orbits in axisymmetric equilibrium magnetic field in GMEC code.

The next step is to verify the multi-particle orbits when considering a perturbed electromagnetic field. In this situation, the toroidal symmetry is broken, $E$ and $P_{\phi}$ are no longer conserved. However, given the perturbed electric potential of
\begin{flalign}\label{eq34}
\begin{split}
    \delta\varphi=A_{\varphi} \exp\left[-\left(\frac{\psi-\psi_0}{\delta\psi}\right)^2\right] \sin(n\phi-m\theta-\omega t),
\end{split}
\end{flalign}
and the perturbed parallel magnetic potential satisfies $E_{\parallel}=-\partial \delta A_{\parallel}/\partial t -\bs{b}\cdot\nabla\delta\varphi=0$, where $A_{\varphi}$ is the amplitude (do not confused with the symbol of vector potential), $\psi_0,\delta\psi$ are the center and width of the perturbed field respectively, $m$ is the poloidal mode number, $\omega$ is the frequency, $E-P_{\phi}\omega/n$ is conserved for each particle. This conservation has also been carefully checked.

\begin{figure}[htbp]
\centering
\includegraphics[width=0.45\textwidth]{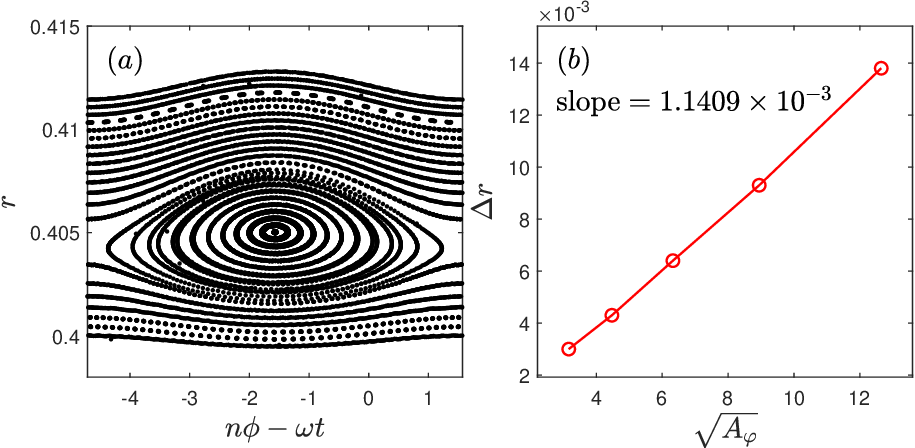}
\caption{(a) Poincaré plot of $p=-1$ resonance, $A_{\varphi}=10\mathrm{V}$, (b) the width of the resonant island in phase space as a function of the square root of mode amplitude.}
\label{fig7}
\end{figure}

The condition that particles resonantly interact with the perturbed electromagnetic field is
\begin{flalign}\label{eq35}
\begin{split}
    \omega=k_{\parallel,p}v_{\parallel}\approx \frac{v_{\parallel}}{R} \left(n-\frac{m+p}{q}\right), p\in \mathbb{Z}.
\end{split}
\end{flalign}
In Figure \ref{fig7}(a), a Poincaré plot of $p=-1$ resonance is illustrated under the parameters of $\psi_0=0.5$, $\delta\psi=0.1$, $\omega=v_{A0}/3R_0$, $n=1$, $m=2$, $A_{\varphi}=10\mathrm{V}$. we can see that there is a resonant island in the phase space. According to the Berk-Breizman theory \cite{Berk19902246}, the width of the resonant island is proportional to the square root of mode amplitude, $\Delta r=a\sqrt{A_{\varphi}}$, where $a$ is the slope and is given analytically by the following formula \cite{Berk19902246},
\begin{flalign}\label{eq36}
\begin{split}
    a=\left(\frac{v_{\parallel}}{B_0\Omega_0}\frac{1}{\omega}\frac{q}{\hat{s}}\frac{m}{r}\right)^{1/2}.
\end{split}
\end{flalign}
Substituting the equilibrium parameters into equation \ref{eq36}, we obtain $a=1.2558\times10^{-3}$, which is consistent with the slope calculated from the numerical solution in Figure \ref{fig7}(b).

\subsection{Simulations of $n=3$ TAE}\label{sec3.2}
Linear simulations of $n=3$ TAE driven by EPs without FLR effect are carried out and results are presented in this section. The equilibrium parameters are selected as $R_0=3\mathrm{m}$, $a_0=1\mathrm{m}$, $B_0=1\mathrm{T}$, the simulation region is $\psi\in [0.01,1]$, density profile is assumed to be uniform with central value of $n_0=10^{19} \mathrm{m}^{-3}$, both bulk plasma and EPs are composed of hydrogen ions, while the EP charge is artificially set to be $2e$, where $e$ is the unit charge. The $q$ profile is set to be $q=1.6667+0.5\psi+0.8333\psi^2$, equilibrium bulk plasma pressure $P_b$ is chosen to be zero. EP slowing-down distribution function is $f_0=[cH(v_0-v)/(v^3+v_c^3)]\exp(-\psi /0.25)$, where $c$ is a normalization factor, $H$ is the step function, $v_0=2v_{A0}$ is the maximum velocity, $v_c=0.58 v_0$ is the critical velocity.

For linear simulations, we add a mode filter after every time step and only the mode with a single $n$ is chosen. The toroidal periodicity is also exploited such that the computational domain is reduced to $1/n$ torus toroidally. As a result, we can use relatively fewer grids in $z$ direction to accelerate computation. In the following simulations, grid numbers are all chosen to be $n_x=256,n_y=64,n_z=16$, and the number of EP markers is $N_p=5\times10^6$.

GMEC simulations in this work are carried out on the Tianhe No.3 high-performance clusters. In each case run, 32 CPUs with 28 threads each are used in parallel simulations. The consuming time for one case of $2\times10^4$ time steps is around 0.4 hours when solving MHD and DK equations, while is around 3.2 hours when solving MHD and GK equations. The speedup is around 20 comparing with the consuming time when using 1 CPU with 28 threads. Further accelerations by optimizing memory access of particles and gyro-averaging implementation will be considered in future. The convergence of the numerical results has also been checked, \textit{i.e.}, the growth rate, real frequency and mode structure are almost the same when using larger grid numbers and/or more EP markers.

\begin{figure}[htbp]
\centering
\includegraphics[width=0.45\textwidth]{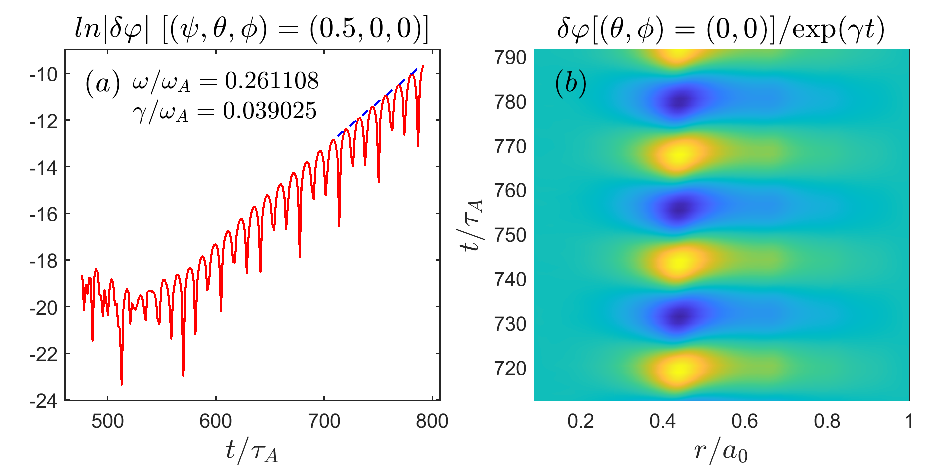}
\caption{Numerical results for EP beta $\beta_h=0.02$. (a) Evolution of $\ln|\delta\varphi|$ at a given spatial position $(\psi,\theta,\phi)=(0.5,0,0)$, (b) Evolution of $\delta\varphi/\exp(\gamma t)$ at $(\theta,\phi)=(0,0)$.}
\label{fig8}
\end{figure}

\begin{figure}[htbp]
\centering
\includegraphics[width=0.45\textwidth]{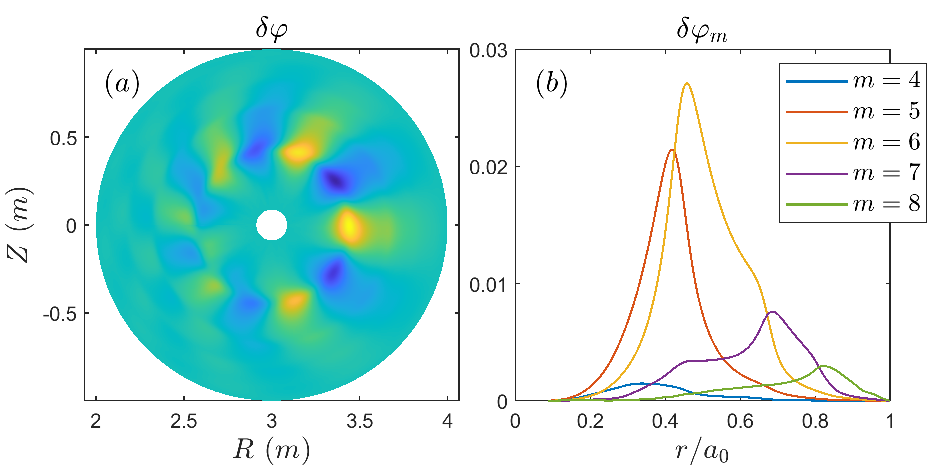}
\caption{Numerical results for EP beta $\beta_h=0.02$. (a) 2D mode structure of $\delta\varphi$ in the poloidal plane of $\phi=0$, (b) Fourier harmonics of different poloidal numbers.}
\label{fig9}
\end{figure}

\begin{figure}[htbp]
\centering
\includegraphics[width=0.4\textwidth]{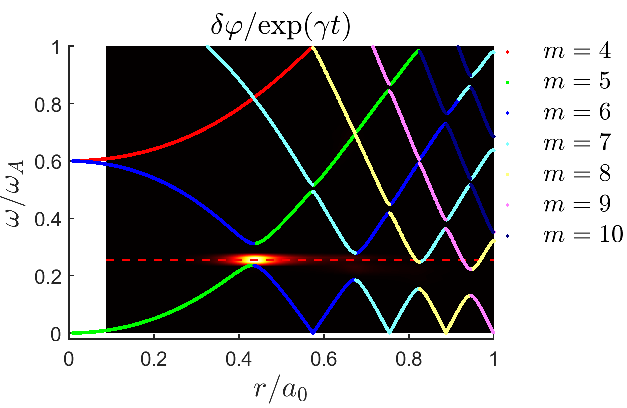}
\caption{The power spectrum of $\delta\varphi/\exp(\gamma t)$ and its location relative to the Alfvén continuum, the red dashed line represents the value of $\omega/\omega_A$ calculated in Figure \ref{fig8}(a).}
\label{fig10}
\end{figure}

Numerical results for EP beta $\beta_h=0.02$ are displayed in figures \ref{fig8}-\ref{fig10}, in which the analytical equilibrium of concentric circles in section \ref{sec2.4} is adopted. The time evolution of $\ln|\delta\varphi|$ at a given spatial position $(\psi,\theta,\phi)=(0.5,0,0)$ is displayed in Figure \ref{fig8}(a), the growth rate and real frequency can be obtained via fitting the peak positions as shown by the blue dashed line. Figure \ref{fig8}(b) shows the spatial-temporal evolution of $\delta\varphi/\exp(\gamma t)$ at a given cross section $(\theta,\phi)=(0,0)$, we can see that the mode is located at $r/a_0 \sim 0.45$, and the oscillating frequencies are the same for all spatial positions, which indicates that it is indeed an eigenmode.

The 2D mode structure of $\delta\varphi$ in $R,Z$ plane and its Fourier decomposition of different poloidal numbers $m$ are displayed in Figure \ref{fig9}. The TAE mode is dominated by $m=5$ and $m=6$ components. The value of $q$ at $r/a_0\sim 0.45$ is around $(5+0.5)/3$, which is consistent with the TAE theory \cite{Cheng1986}.

The power spectrum of $\delta\varphi/\exp(\gamma t)$ at the mid-plane of $(\theta,\phi)=(0,0)$ is displayed in Figure \ref{fig10}, and it is located just above the tip of TAE gap of $m=5$ and $m=6$, where the Alfvén continuum is obtained from the ALCON code \cite{Deng2012}. The red dashed line represents the value of $\omega/\omega_A$ calculated in Figure \ref{fig8}(a). This result further demonstrates that the mode is indeed a TAE.

\begin{figure}[htbp]
\centering
\includegraphics[width=0.45\textwidth]{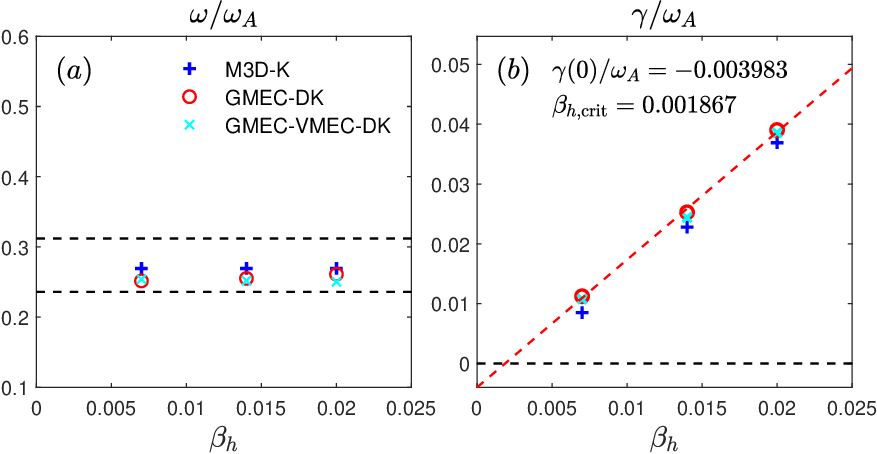}
\caption{(a) Real frequency and (b) growth rate of M3D-K (blue pluses), GMEC with analytical equilibrium (red circles) and GMEC with VMEC numerical equilibrium (cyan crosses).}
\label{fig11}
\end{figure}

The VMEC equilibrium has also been adopted in the GMEC simulations, and the corresponding the real frequency and growth rate are almost the same as the result of analytical equilibrium as shown in Figure \ref{fig11}. These results are in good agreement with those of M3D-K code \cite{Fu2006}.

\subsection{Benchmarks of the $n=6$ TAE}\label{sec3.3}
A widely used benchmark case of the $n=6$ TAE with various eigenvalue, kinetic and hybrid codes has been reported in [A. Könies \textit{et al.}, Nuclear Fusion \textbf{58}, 126027 (2018)] \cite{Könies2018}. The equilibrium parameters are selected as $R_0=10\mathrm{m}$, $a_0=1\mathrm{m}$, $B_0=3\mathrm{T}$, the bulk ions are hydrogen with a flat density profile of $n_0=2\times 10^{19} \mathrm{m}^{-3}$. The $q$ profile is set to be $q(r)=1.71+0.16(r/a_0 )^2$, bulk plasma pressure is $P_b (s)=(7.17\cdot 10^3-6.811\cdot 10^3 s-3.585\cdot 10^2 s^2 )\mathrm{Pa}$, EP (deuterons) density profile is $n(s)=n_0c_3\exp(-c_2/c_1\tanh((\sqrt{s}-c_0)/c_2)$, where $n_0=1.44131\cdot 10^{17} \mathrm{m}^{-3}$, $c_0=0.49123$, $c_1=0.298228$, $c_2=0.198739$, $c_3=0.521298$, $s=\psi_t/\psi_{tm}$, $\psi_t$ is the toroidal magnetic flux, $\psi_{tm}$ is the boundary value of $\psi_t$. The EP distribution function in velocity space is taken to be a Maxwellian with temperature ranging from $100\mathrm{keV}$ to $800\mathrm{keV}$. The simulation region is set to be $\psi\in [0.01,1]$. VMEC numerical equilibrium is obtained with $q$ and $P_b$ profiles as input.

\begin{figure}[htbp]
\centering
\includegraphics[width=0.45\textwidth]{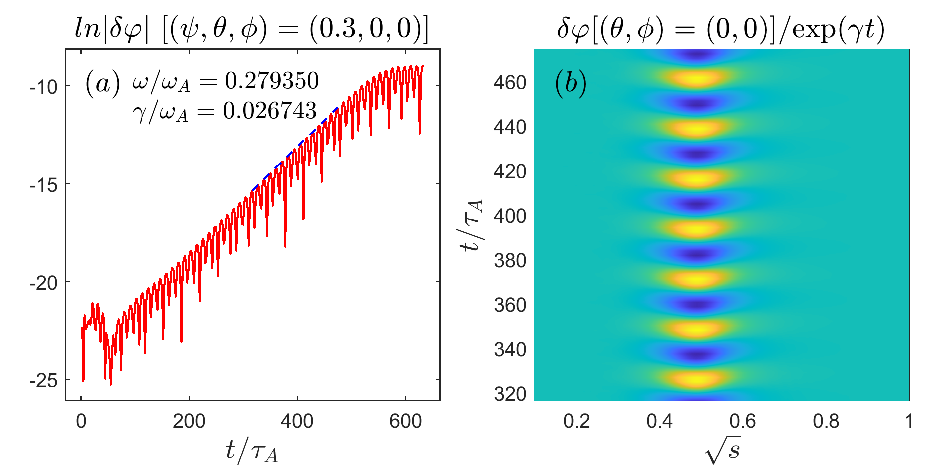}
\caption{Numerical results for EP temperature $T_f=400\mathrm{keV}$. (a) Evolution of $\ln|\delta\varphi|$ at a given spatial position $(\psi,\theta,\phi)=(0.3,0,0)$, (b) Evolution of $\delta\varphi/\exp(\gamma t)$ at $(\theta,\phi)=(0,0)$.}
\label{fig12}
\end{figure}

\begin{figure}[htbp]
\centering
\includegraphics[width=0.45\textwidth]{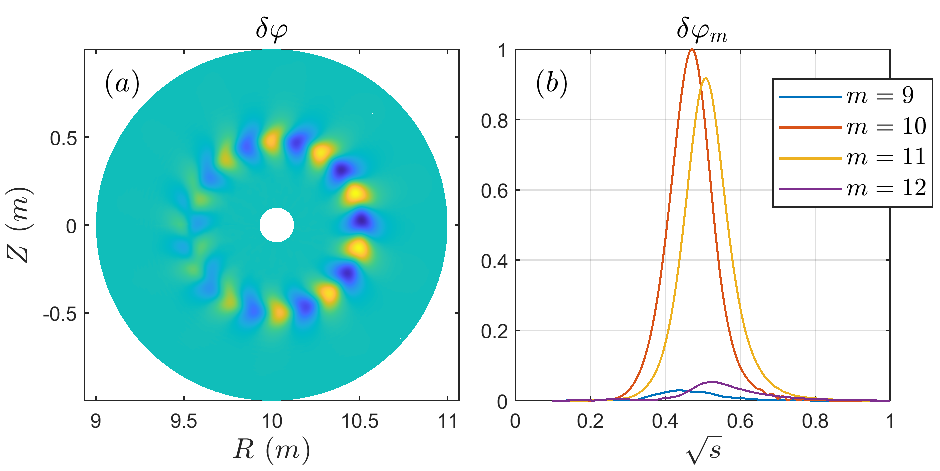}
\caption{Numerical results for EPs temperature $T_f=400\mathrm{keV}$. (a) 2D mode structure of $\delta\varphi$ in poloidal plane of $\phi=0$, (b) Fourier harmonics of different poloidal numbers.}
\label{fig13}
\end{figure}

Numerical results for EP temperature $T_f=400\mathrm{keV}$ without FLR effect are displayed in figures \ref{fig12}-\ref{fig13}. The evolution of $\ln|\delta\varphi|$ at a given spatial position is displayed in Figure \ref{fig12}(a), the growth rate and real frequency can be calculated via fitting the peak positions as shown by the blue dashed line. Notice that the mode saturates after $550\tau_A$ due to particle nonlinearity. The spatial-temporal evolution of $\delta\varphi/\exp(\gamma t)$ at the mid-plane is displayed in Figure \ref{fig12}(b). The eigenmode peaks around $\sqrt{s}\sim 0.5$.

The 2D mode structure of $\delta\varphi$ in $R,Z$ plane and also its Fourier decomposition of different poloidal numbers $m$ are displayed in Figure \ref{fig13}. The TAE mode is dominated by $m=10$ and $m=11$. The value of $q$ at $\sqrt{s}\sim 0.5$ is around $(10+0.5)/6$, which is consistent with the TAE theory \cite{Cheng1986}. The simulated mode structure agrees well with the results of other codes \cite{Könies2018} and M3D-C1-K \cite{Liu2022}.

\begin{figure}[htbp]
\centering
\includegraphics[width=0.45\textwidth]{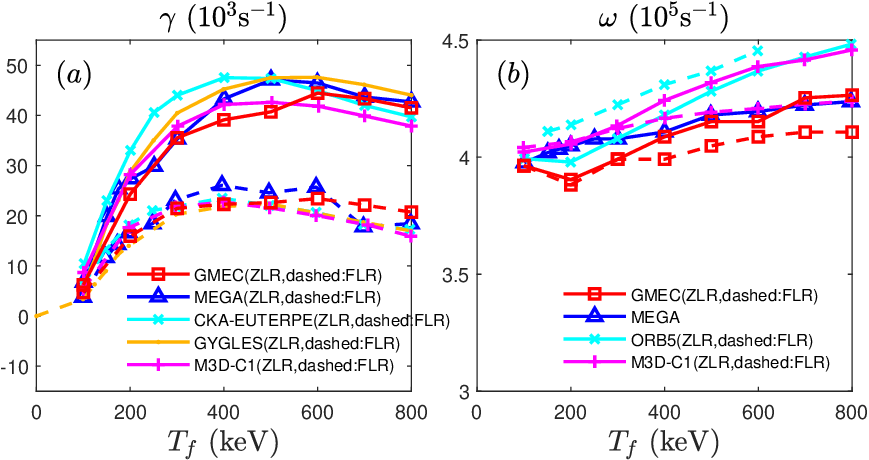}
\caption{(a) The growth rate and (b) real frequency as function of EP temperature from results of ZLR (solid lines) and FLR (dashed lines) of GMEC and several other codes. Results of other codes are copied from Fig.5 in [C. Liu \textit{et al.}, Computer Physics Communications \textbf{275}, 108313 (2022)] \cite{Liu2022}.}
\label{fig14}
\end{figure}

The growth rate and real frequency as function of EP temperature are displayed in Figure \ref{fig14}, including results without FLR effect (also named ZLR) and with FLR effect. The growth rates are close to the results from other GK, hybrid and eigenvalue codes \cite{Könies2018}, and M3D-C1-K \cite{Liu2022}. The real frequencies of ZLR are close to the results of MEGA \cite{Könies2018}, while the real frequencies with FLR effect are reasonably close to the results of MEGA.

\subsection{Benchmarks of RSAE in DIII-D experiments}\label{sec3.4}
We also performed linear simulations of RSAE in DIII-D experiments \cite{Collins2016} with FLR effect. In these simulations, real tokamak geometry with plasma equilibrium and EP distribution from experimental diagnostics are taken into account. The equilibrium parameters follow the setup in [S. Taimourzadeh \textit{et al.}, Nuclear Fusion \textbf{59}, 066006 (2019)] \cite{Taimourzadeh2019}, in which various eigenvalue, GK and hybrid codes have participated in a linear benchmark. The $q$ profile has a minimum $q_{\mathrm{min}}=2.94$ at $\rho=0.4$, where $\rho$ is the normalized square root of toroidal magnetic flux. The EP distribution in velocity space is approximated to be an isotropic Maxwellian, and EP density and temperature profiles are measured from kinetic EFIT. The VMEC numerical equilibrium is adopted with $q$ and the total pressure profiles as input. The shape of the last closed flux surface is described by the analytical expressions of $R=R_0+r \cos[\theta_s + (\mathrm{sin}^{-1}\bar{\delta})\sin\theta_s]$, 
$Z=\bar{\kappa}r \sin\theta_s$, where $\bar{\kappa}=1.49557$ and $\bar{\delta}=0.0528165$ are the averaged elongation and triangularity.

\begin{figure}[htbp]
\centering
\includegraphics[width=0.45\textwidth]{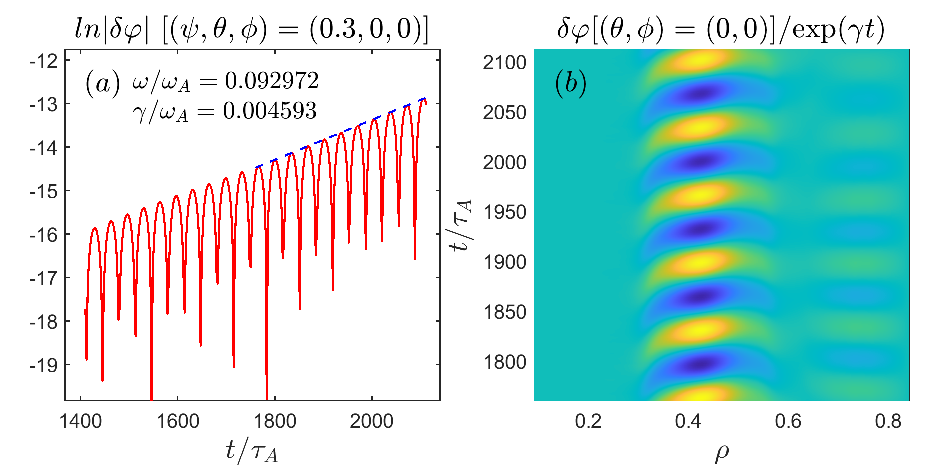}
\caption{Numerical results of $n=4$ RSAE. (a) Evolution of $\ln|\delta\varphi|$ at a given spatial position $(\psi,\theta,\phi)=(0.3,0,0)$, (b) Evolution of $\delta\varphi/\exp(\gamma t)$ at $(\theta,\phi)=(0,0)$.}
\label{fig15}
\end{figure}

\begin{figure}[htbp]
\centering
\includegraphics[width=0.45\textwidth]{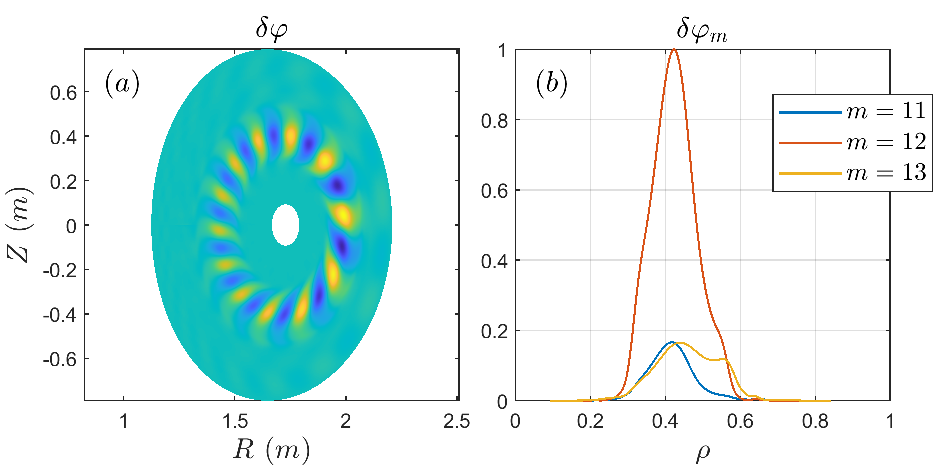}
\caption{Numerical results of $n=4$ RSAE. (a) 2D mode structure of $\delta\varphi$ in poloidal plane of $\phi=0$, (b) Fourier harmnics of different poloidal numbers.}
\label{fig16}
\end{figure}

Numerical results of the $n=4$ RSAE with FLR effect are displayed in figures \ref{fig15}-\ref{fig16}. The time evolution of $\ln|\delta\varphi|$ at a given spatial position is displayed in Figure \ref{fig15}(a). Figure \ref{fig15}(b) shows the spatial-temporal evolution of $\delta\varphi/\exp(\gamma t)$ at a given cross section. The mode is located around the position of $q_{\mathrm{min}}$.

The 2D mode structure of $\delta\varphi$ in $R,Z$ plane and its Fourier decomposition of different poloidal numbers $m$ are displayed in Figure \ref{fig16}. The RSAE mode is located around $\rho=0.4$, and is dominated by $m=12$ and is weakly coupled to $m=11,13$. The simulated mode structure agrees well with the results of EUTERPE and GTC \cite{Taimourzadeh2019}.

\begin{figure}[htbp]
\centering
\includegraphics[width=0.3\textwidth]{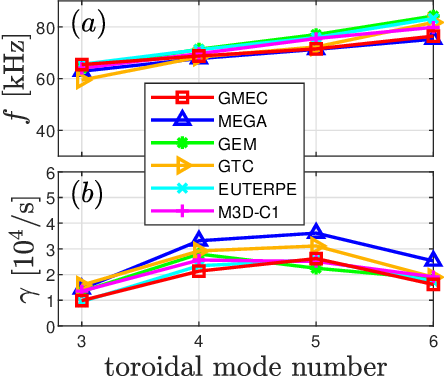}
\caption{(a) The growth rate and (b) real frequency as function of $n$ from GMEC and other codes. Results of other codes are copied from Fig.7 in [C. Liu \textit{et al.}, Computer Physics Communications \textbf{275}, 108313 (2022)] \cite{Liu2022}.}
\label{fig17}
\end{figure}

The growth rate and real frequency as function of $n$ are displayed in Figure \ref{fig17}. The linear dispersion of $n=3-6$ from GMEC agrees well with results of the multi-code verification and validation simulations \cite{Taimourzadeh2019, Liu2022}. The real frequency increases with growing $n$, while the most unstable modes are $n=4$ and $n=5$.

\section{Summary}\label{sec4}
In summary, we have developed a new hybrid code GMEC that can numerically simulate energetic particle-driven Alfvén instabilities and energetic particle transport in fusion plasmas. Up to now, a simplified version of the hybrid code has been completed with several verifications and benchmarks.

In GMEC code, the bulk plasma is described by the reduced MHD model of equation \ref{eq1}. As for energetic particles, the gyrokinetic equations are solved via PIC method and the perturbed distribution function of energetic particles is solved by using $\delta f$ method and is used to compute the perturbed energetic particle pressures that are coupled into the reduced MHD equations. The field-aligned coordinates and meshes are adopted to resolve Alfvén eigenmodes with high toroidal numbers effectively. For spatial discretization, five-points finite difference method up to fourth order precision has been adopted. For time advance, we use fourth order Runge-Kutta method both for the reduced MHD equations and gyrokinetic equations of motion.

GMEC is parallelized using multi-process with MPI and multi-threading with TBB library. Both analytical and VMEC numerical equilibria are supported. Details of the MHD equation solver and benchmarks of MHD modes are presented in Part I \cite{jiang2024development} of series paper. This paper, the part II of the series paper, emphasizes the gyrokinetic equation solver and linear simulations of energetic particle-driven Alfvén eigenmodes. Single particle orbits in axisymmetric tokamak equilibrium and multi-particle orbits in the presence of a perturbed electromagnetic field have been successfully verified. Three benchmark cases of AEs are presented, including (1) the $n=3$ TAE without FLR effect with both analytical and VMEC numerical equilibrium: the GMEC simulation results agree very well with results of M3D-K, (2) the $n=6$ TAE with and without FLR effect: the growth rate and mode structure from GMEC are in good agreement with the results of several other codes \cite{Könies2018}, (3) the $n=3-6$ RSAEs with FLR effect: results are close to those of the multi-code verification and validation simulations \cite{Taimourzadeh2019}.

For the future work, the fluid nonlinearity and high-order FLR effects will be retained in the extended MHD equations. Furthermore, kinetic effects of thermal ions will be included by calculating the thermal ion pressures kinetically. Finally, the efficiency of PIC part will be improved by optimizing memory access of PIC simulation.

\section*{Acknowledgement}
The authors are indebted to Dr. Chang Liu who provided the equilibrium of DIII-D for RSAE benchmarks. One of the authors (Zhaoyang Liu) would like to thank Dr. Yao Zhou, Dr. Zhicheng Feng and Dr. Guodong Yu for helpful discussions. GMEC runs in this work are carried out on Tianhe No.3 high-performance clusters (https://www.nscc-tj.cn/cjjs\_zy\_th3). This work is supported by the National MCF Energy R\&D Program of China (Grant No. 2019YFE03050001) and Zhejiang Lab under the project of Digital twin system for intelligent simulation and control of nuclear fusion (124000-AC2304).

\nocite{*}
\bibliography{ref}

\end{document}